\documentclass[12pt]{article}
\usepackage[T1]{fontenc}
\usepackage{amsfonts}
\usepackage{amsmath,amsthm, amssymb}
\usepackage{authblk, enumerate, natbib}
\usepackage{multicol}
\usepackage{graphicx} 
\usepackage{mathtools}

\usepackage{algorithm}
\usepackage{algpseudocode}

\usepackage{geometry}
\newtheoremstyle{mytheoremstyle} % name
    {\topsep}          % Space above
    {\topsep}          % Space below
    {\itshape\fontfamily{ptm}\selectfont}          % Body font
    {}              % Indent amount
    {\fontfamily{ptm}\selectfont\scshape}          % Theorem head font
    {.}             % Punctuation after theorem head
    {.5em}            % Space after theorem head
    {} % Theorem head spec (can be left empty, meaning ‘normal’)
\theoremstyle{mytheoremstyle}
\newtheorem{theorem}{Theorem}
\newtheorem{lemma}{Lemma}
\theoremstyle{remark}

\newtheorem{example}{Example}

\def\mod{\operatorname{mod}}
\def\0{{\mathbf 0}}

\def\1{{\mathbf 1}}

\let\oldbibliography\thebibliography
\renewcommand{\thebibliography}[1]{\oldbibliography{#1}
\setlength{\itemsep}{6pt}} %Reducing spacing in the bibliography.

\title{$Q_B$-Optimal Two-Level Designs}
\author[1]{Pi-Wen Tsai}
\affil[1]{National Taiwan Normal University}
\author[2]{Steven G.\ Gilmour}
\affil[2]{King's College London}
\date{}

\begin{document}

%%%%%%%%%%%%%%%%%%%%%%%%%%%%%%%%%%%%%%%%%%%%%%%%%%%%%%%%%%%%%%%%%%%%%%%%%%%%%%%%%%%%%%%%%%%%%%%%%%%%%%%%%%%%%%%%%%%%%%%%%%%%

\fontsize{12}{14pt plus.8pt minus .6pt}\selectfont \vspace{0.8pc}
\centerline{\large\bf $Q_B$-Optimal Two-Level Designs }
%\vspace{2pt} 
%\centerline{\large\bf HERE IF A SECOND LINE IS NEEDED}
\vspace{.4cm} 
\centerline{Pi-Wen Tsai and Steven G.\ Gilmour} 
\vspace{.4cm} 
\centerline{\it National Taiwan Normal University and King's College London}
 \vspace{.55cm} \fontsize{9}{11.5pt plus.8pt minus.6pt}\selectfont

%%%%%%%%%%%%%%%%%%%%%%%%%%%%%%%%%%%%%%%%%%%%%%%%%%%%%%%%%%%%%%%%%%%%%%%%%%%%%%%%%%%%%%%%%%%%%%%%%%%%%%%%%%%%%%%%%%%%%%%%%%%%

\begin{quotation}
\noindent {\it Abstract:}
%\begin{abstract}
Two-level designs are widely used for screening experiments where the
goal is to identify a few active factors which have major effects.
Orthogonal two-level designs in which all factors are level-balance
and each of the four level combinations of any pair of factors appears
equally often are commonly used.
In this paper, we apply the model-robust $Q_B$ criterion introduced by
Tsai, Gilmour and Mead (2007) to the selection of optimal two-level
screening designs without the requirements of level-balance and
pairwise orthogonality.  The criterion  incorporates experimenter's 
prior belief on how likely a factor is to be active
and recommends different designs under different priors, and without
the requirement of level-balance and
pairwise orthogonality, a wider range of designs is possible. 
A coordinate exchange algorithm is developed for the construction of
$Q_B$-optimal designs for given priors.

\vspace{9pt}
\noindent {\it Key words and phrases:}
%\noindent
%\emph{keyword: }
$Q_B$-criterion, $G_2$-Aberration; 
E($s^2$); UE($s^2$); Nonorthogonal design; 
Generalized word count; 
Coordinate Exchange; Model Uncertainty 
%\end{abstract}

\par
\end{quotation}\par

\def\thefigure{\arabic{figure}}
\def\thetable{\arabic{table}}

\renewcommand{\theequation}{\thesection.\arabic{equation}}

\fontsize{12}{14pt plus.8pt minus .6pt}\selectfont

\section{Introduction}

%Highly fractionated regular or irregular two-level fractional
%factorial designs are popular for screening experiments where

The goal of screening experiment is to identify a few active effects
among many, making the effect sparsity assumption.  Therefore it is
natural to use two-level orthogonal designs so that each of the main
effects can be estimated with maximum precision and independently from
other main effects.  Highly fractionated regular fractional factorial
designs which are
%is a carefully chosen subset of treatment combinations, 
determined by some defining words and factorial effects are either
orthogonal to or completely aliased with each other are commonly
used. These are orthogonal main effects plans where each of the four
level combinations of any pair of factors has the same number of
occurrences and all factors are level-balanced.  Regular factorial
designs, however, exist only when $N$, the number of runs, is a power
of 2.  Irregular factorial designs, such as Plackett-Burman designs
where at least one pair of effects is neither completely orthogonal
nor totally aliased, are popular for their run-size flexibility. These
have more complex aliasing structures among effects than regular
designs. A popular criterion for choosing two-level regular or
irregular designs is minimum aberration (\cite{fries80}) or
generalized minimum aberration (\cite{tang99}). The aberration
criterion was originally defined from the combinatorial point of view
based on the \emph{effect hierarchy assumption} that lower order
effects are more important than higher order effects and effects of
the same order are equally important.  These aberration-type criteria
concentrate first on minimizing aliasing between pairs of main
effects, then on minimizing aliasing between main effects and
two-factor interactions, then on minimizing aliasing between pairs of
two-factor interactions and so on.

Tsai, Gilmour and Mead (2007) went beyond the traditional approach by
suggesting the model-robust $Q_B$ criterion which incorporates
experimenters' prior knowledge on the probability of each effect being
in the best model.  The use of the $Q_B$ criterion requires a
definition of the maximal model of interest and assumes one of
submodels of the maximal model will be the best model.  The
$Q_B$-criterion is defined as the weighted average of the approximation of
the $A_s$-efficiency (excluding the intercept) for each of the possible submodels, with weight depending on the prior probability of the model being the best model.
Like most work in the design literature,
in \cite{tsai07} the $Q_B$ criterion is used as a secondary criterion
among the class of level-balanced or orthogonal main effects designs. \cite{tsai10} 
showed that for the first-order maximal main effect model, the $Q_B$
criterion is to select a design by miminising a linear combination of
the aliasing between main effects and the intercept and the aliasing
between pairs of main effects. However, in their example they use the $Q_B$ criterion to select a design
that minimizes pairwise orthogonality among the class of
level-balanced designs in which the two levels appear the same number of times. This approach is equivalent to the standard
approach for supersaturated designs where the $E(s^2)$-criterion is used
among the class of designs with all factors level-balanced. 

In this
paper, we use the $Q_B$ criterion as a primary criterion and focus on
the application of two-level screening designs without the
requirements of level-balance and pairwise orthogonality. 
Additionally a coordinate exchange algorithm is developed to
generate $Q_B$-optimal designs without the requirements of
level-balance or pairwise orthogonality.
Applications of the $Q_B$-criterion to the first-order maximal model with supersaturated, saturated and unsaturated screening designs and to the second-order maximal model 
are given. 
The algorithm generates a wider range of $Q_B$-optimal first-order
designs which respect experimenter's prior belief on the importance of
a factor and the explicit trade-off between the level-balance and
pairwise orthogonality are demonstrated. In general, a wide range of
two-level $Q_B$-optimal designs that would jointly minimise the
aliasing among different orders of factorial effects are generated.
% a coordinate exchange algorithmic is developed to generate the
%$Q_B$-optimal designs without the requirements of level-balance or
%pairwise orthogonality.  , a wider range of two-level screening
%designs can be found. In this paper, A wide range of two-level
%designs which reflexing experimenter's prior belief on the importance
%of each effect would be found.
Recently, \cite{vazquez23} provide efficient algorithms for generating two-level $Q_B$-optimal designs, using exact and heuristic methods. They focus on the computational strategies for the construction of $Q_B$-optimal designs and in this paper, we emphasise the various applications of 
two-level experiments.

This paper is organized as follows.  The definition of the $Q_B$-criterion
is reviewed in Section 2 along with the notation for summarising the
aliasing among different orders of factorial effects.  The coordinate
exchange algorithm is discussed in Section 3.  The applications of
$Q_B$-optimal designs for the first-order maximal model and the
second-order maximal model are given in Sections 4 and 5. Some
concluding remarks are made in Section 6.

%Tsai, Gilmour and Mead (2007) went beyond this traditional approach
%by suggesting the model-robust $Q_B$ criterion by incorporating
%experimenters' prior knowledge on the importances of each factor.
%Like most work in design literature, the designs discussed in
%\cite{tsai10} focus on level-balanced designs (with each column of
%$D$ has the same number of $\pm 1$) or orthogonal main effects
%designs (with every pair of columns of $D$ has the same number of
%[$\pm 1, \pm 1$]) and use $Q_B$ criterion as the secondary
%criterion. For example, In the context of supersaturated designs with
%$m\ge N$ and the first-order main effect model is the maximal model
%of interest, the $Q_B$ criterion is to select a design by miminising
%a linear combination of measures of level-balance and pairwise
%orthogonality. However, instead of finding a $Q_B$-optimal criterion
%under different priors, \cite{tsai10} use the $Q_B$ criterion to
%select a design that minimizes pairwise orthogonality among the class
%of level-balanced designs which is equivalent to the standard
%$E(s^2)$-criterion for supersaturated designs.  Thus in this paper,
%we will demonstrate different cases with using $Q_B$ criterion as the
%primary objective and try to find $Q_B$-optimal designs without the
%requirements of level-balanced or pairwise orthogonality.

\section{The $Q_B$-criterion}

For an $N$-run design with $m$ two-level factors, let $y$ be the
response variable and $y=X\beta+\varepsilon$ be the maximal model of
interest where $\beta=[\beta_0, \beta_1, \cdots, \beta_v]^{\top}$ is
the $(v+1)\times 1$ vector of parameters in the maximal model and $X$
is the corresponding model matrix.  Notice that the form of maximal
model could be first-order model, second-order model, or
higher-order.  The maximal model is not required to be estimable
and is often determined by the combination of $N$ and $m$.  It is
assumed that one of the submodels of the maximal model will be the
final model that we will end up fitting. The $Q_B$-criterion is
defined as the weighted average of the approximations of the variances of
the parameter estimators of $\beta_1, \cdots, \beta_v$ (excluding
$\beta_0$) in each of the possible submodels, with weight depending on
the prior probability of the model being the best model.  Letting
$(a_{ij})$, $i, j=0, 1, \cdots v$, be the element of $X^\top X$,
\cite{tsai07} derived that
\begin{equation}
 Q_B=
 %\sum_{t=1}^{||\Delta||} w_t
 %\tilde{A}_s(\mathcal{M}_t)=%\sum_{i=1}^{v}
 %\frac{1}{a_{i0}}\,\frac{a_{i0}^2}{a_{ii}a_{00}}\ p_{i0}+
 \sum_{i=1}^{v}\sum_{j=0}^{v} \frac{1}{a_{ii}}\,
 \frac{a_{ij}^2}{a_{ii}a_{jj}}\ p_{ij}.
\label{eq:QB17}
\end{equation}
Here the intercept $\beta_0$ is treated as a nuisance parameter, the
precision for the estimate of the intercept is not of interest, so the
index of $i$ starts from 1.  But the aliasing of the intercept and a
factorial effect still affects the precision of the estimate of the
factorial effect, so the index $j$ starts from 0.  For an $N$-run design
with $m$ two-level factors, the diagonal elements $a_{ii}=N$, for all
$i$, so we write the $Q_B$-criterion as
\begin{equation}
Q_B=\sum\limits_{i=1}^{v} p_{i0} \left({a^2_{i0}}/{N^2}\right) +
\mathop{\sum\limits_{i=1}^{v}\sum\limits_{j=1}^{v}}_{i\ne j} p_{ij}
\left({a^2_{ij}}/{N^2}\right),
\label{eq:QB}
\end{equation}
where $p_{i0}$ is the cumulative prior sum of the probability of a
model being the best model, where the sum is done over models
containing the factorial effect that $i$ refers to, $i=1, \cdots,
v$; and $p_{ij}$ is the cumulative prior sum of the probability of a
model being the best model, where the sum is done over models
containing both the terms that $i$ and $j$ refer to for $i\ne j$.

The generalization and application of the $Q_B$-criterion to different
types of designs are presented in \cite{tsai10}. They showed that the
$Q_B$-criterion can be used in many different situations, such as
regular or irregular fractional factorial designs with two or three
levels, saturated or unsaturated designs, and it provides a bridge
between alphabetic optimality and aberration.  To study the
$Q_B$-criterion with other commonly used criteria for designs with two
or three levels or mixed levels, \cite{tsai10} introduced the
``generalized word count (GWC)'' which summarises the overall aliasing
for factorial effects of a given number of factors where some factors
are at some particular orders.

For two-level designs, let $X_d=[X_1, \cdots, X_m]$ be the treatment
factors where each factor has entries labeled $-1$ and 1, and
$x_i=[x_{i1}, \cdots, x_{im}]^t$ be the $i$th element of ${X}_d$, $i=1,
\cdots, N$.  For a particular $k$-factor factorial effect $s$, which
is a subset of $\{1, \cdots, m\}$, let $X_s$ be the set of $s$
corresponding columns of $X_d$. Define
\[
R_k(s)=\frac{1}{N^2}\left[ \sum_{i=1}^N ( X_{i,s_{1}} \cdots X_{i,
    s_k}) \right]^2,
\]
where $X_{i,s_j}$ is the $i$th level-combination of the $j$th column
in $X_s$.  $R_k(s)$ is the square of the sum of the
element-by-element products for these $s$ columns divided by $N^2$,
which is a measure of aliasing of the factorial effect $s$ and the
intercept.  For example, let $s=\{1, 2, 5\}$, then $R_3(\{1, 2, 5\})$
is a measure of aliasing for factorial effect $X_1X_2X_5$ and the
intercept.  If the resulting products have the same number of $\pm
1$s, then $R_k(s)=0$ and the factorial effect is orthogonal to the
intercept; if the resulting products are all equal to 1 or all equal
to $-1$, then $R_k(s)=1$ and the factorial effect is fully aliased
with the intercept. In general, $0\le R_k(s)\le 1$ since any factorial
effect might be neither orthogonal to nor fully aliased with the
intercept for a two-level design. This is the same as the
$J$-characteristic for the two-level designs discussed in
\cite{tang01}.

Let 
\begin{equation}
b_k=\sum_{s: |s|=k} R_k(s),
%  = \sum_{s: |s|=k} \Big\{ \frac{1}{N^2 \Big[\sum_{i=1}^N (
%  x_{i1}\times \cdots \times x_{ik}) \Big]^2 \Big\}.
\label{eq:bk}
\end{equation}
which is the sum of $R_k(s)$ for all possible factorial effects
with $k$ factors out of the possible $m$ factors. The vector $b_1,
b_2, b_3, \cdots, b_m$ is the GWC for two-level designs which
summarises the overall aliasing for factorial effects with $k$
two-level factors and the intercept. 
% If $b_k=0$, the all factorial effects of $k$ factors are orthogonal
% to the intercept.
Note that $b_k$ measures not only the overall aliasing between the
$k$-factor interactions and the intercept, it also measures the
overall aliasing of pairs of factorial effects corresponding to two
mutually exclusive partitions of these $k$ factors.  For example,
$b_2=0$ means not only that two-factor interactions are orthogonal to
the intercept, but also every pair of main effects is orthogonal to
each other; $b_3=0$ means that not only that three-factor interactions are
orthogonal to the intercept, but also any main effect is orthogonal to
any two-factor interaction not involving that main effect. 
For two-level designs, the GWC is 
equivalent to the number of defining words in the defining relation for regular two-level designs and is equivalent to the $B_k$ words in the generalized $G_2$ aberration defined in \cite{tang99} for irregular design. 
Based on the effect hierarchy assumption, aliasing among lower-order effects is
less desirable and the aberration-type criteria for regular or irregular
designs are to sequentially minimizing $b_1, b_2, b_3, b_4$ and so
on. The $Q_B$-criterion selects designs by jointly minimizing these
words with the form of the criterion depending on the maximal model of
interest and the weight on each word depending on the prior
information of each effect being in the model.

\section{Coordinate Exchange Algorithm}

One of the most commonly used algorithms to generate optimal
experimental designs is the coordinate-exchange algorithm of
\cite{meyer95}.  Here we proposed an algorithmic coordinate approach
to generate $Q_B$-optimal design. The algorithm can be briefly
described as follows.

%\begin{enumerate}

%\item Starts by generate an $N\times m$ design with labeled $-1$ and
%$1$ and regard each design point as a vector of $m$ elements and then
%calculate the value of $Q_B$-criterion function.

%\item Iteratively examine each element in the design by exchanging
%the value of the coordinate ($-1$ for $1$ or vice versa). If the
%value of $Q_B$-criterion function is lower from making the exchange,
%update the design with the exchanged design and the current criterion
%value becomes the new minimum.

%\item Repeat step 2 until $Q_B$ reaches its lower bounds or no
%improvement is possible.  \end{enumerate}

%The whole cycle is repeated several times with different random
%starting designs and the one with the lowest value is the best design
%found.

For a given prior, we compute the $Q_B$-criterion value of a random
design.  Then the algorithm tries to improve the design by switching
the signs of each of its coordinates in a systematic way. If a sign
switch in a coordinate improves the value of the $Q_B$-criterion, we
update the design and go back to switching signs in the newly best
design. The algorithm stops when the improvement of the criterion
value is less than a small value $\epsilon$ or the number of
iterations equal to the maximum number of iterations $T$. The
pseudo code for this procedure is given in Algorithm~\ref{alg:QBmain}.

\begin{footnotesize}
\begin{algorithm}
\caption{coordinate exchange algorithm }\label{alg:QBmain}
\begin{algorithmic}[1]
  \Require
  Number of runs $N$; number of two-level factors $m$; prior
  probability of each effect being in the model  
  \Require  the maximal number of iterations $T$, a small value $\epsilon$
  \State Initialization:  a random starting design $d$;   $qb_0\gets Q_B(d)$
  \State iter $\gets $ 0; dff $\gets $ a large number
  \While{dff $>\epsilon$ and iter $< T$}
  \For{$i \gets $ 1 to  N}
  \For{$j \gets $ 1 to  m}
  \State Sign switch for  $(i,j)$th coordinate of $d$  to $d^*$; $qb\gets Q_B(d^*)$
  \If{improve, i.e., $qb <qb_0$}
  \State $d \gets  d^*$; $qb_0\gets qb$; dff $\gets  qb_0-qb$; iter $\gets  0$
  \ElsIf{no improve}
  \State iter $\gets  $ iter + 1
  \EndIf
  \EndFor
  \EndFor
\EndWhile
\State Return the best design $d$ 
\end{algorithmic}
\end{algorithm}
\end{footnotesize}

This is a local search algorithm, and to avoid getting stuck at a local best
design, we restart the procedure with different random initial
designs. The coordinate exchange algorithm is not guaranteed to find
the optimal design, but it usually can find designs which are either
optimal or very close to being optimal. Coordinate exchange can
struggle especially when orthogonal main effects designs, or other
designs with a very specific combinatorial structure, are
optimal. Hence, it is usually worthwhile comparing such designs with
those obtained from coordinate exchange, to check that they are
suboptimal, as well as to see how much we lose in terms of $Q_B$
efficiency by insisting on orthogonality and/or level-balance.

\section{First-order maximal model}
For the first-order main effects maximal model, the maximal model is
$E(y)=\beta_0+\beta_1 x_1+\cdots +\beta_m x_m$, with $v=m$.  Then the
model matrix is $X=[1\; X_d]$ where $X_d$ is the $N\times m$ design
matrix.  Using the GWC defined in \eqref{eq:bk}, we have
$\sum_{i=1}^{m} a_{i0}^2/N^2$ equal to $b_{1}$ and $\sum\sum_{i\ne
  j} a_{ij}^2/N^2$ equal to $2b_{2}$.  Assume all factors are
exchangeable and each factor has the same prior probability of being
in the best model. 
Let $\pi_1$ denote the prior probability that a
main effect of a factor is in the best model; then the prior
probability for a model containing main effects of a given $a$ factors
being the best is $\pi_1^a(1-\pi_1)^{m-a}$.
Under the exchangeability assumption, the prior sum for models
containing $X_1$ is the same as that for models containing $X_2$, 
i.e.\ $p_{10}$ in equation \eqref{eq:QB} is the same as $p_{20}$, and
similarly, the prior sum for models containing the pair $X_1$ and $X_2$ 
is the same as that for models containing the pair $X_1$ and $X_3$, i.e.\ $p_{12}=p_{13}$. 
Let $p_{i0}=\xi_1$ for all $i$, and $p_{ij}=\xi_2$ for all $i\ne j$ denote two such prior sums; then we have
$\xi_1=\pi_1\left( \sum\limits_{a=0}^{m-1}
\pi_1^a(1-\pi_1)^{m-1-a}\right)=\pi_1$ where the sum in the brackets
is 1, and $\xi_2=\pi^2_1\left( \sum\limits_{a=0}^{m-2}
\pi_1^a(1-\pi_1)^{m-2-a}\right)=\pi_1^2$.  

Putting the above results
together, the $Q_B$-criterion for the first-order model is to select a
design that minimises
 \begin{equation}
Q_B=\pi_1 b_1+2 \pi_1^2 b_2,
 \label{eq:QB.1st}
\end{equation}
which is a weighted average of the measures of level-balance $(b_1)$
and pairwise orthogonality $(b_2)$. When $\pi_1$ is small and
approaches 0, then $\pi_1^2$ is even smaller and can be negligible. In
this case $b_1$ plays a more important role in the criterion, so
designs with more level-balanced factors and smaller values of $b_1$
tend to be $Q_B$-optimal. When $\pi_1$ is large and approaches 1, then
$b_2$ is almost as important as $b_1$ and we might need to relax the
requirement of level balance in order to have designs where the
aliasing between pairs of main effects is less serious. In other
words, designs with more level-balanced factors are recommended when
the expected number of active factors is small, but when the expected
number of active factors is higher, designs with some
non-level-balanced factors but less serious pairwise aliasing might be
recommended.  Thus the criterion provides an explicit relation for
the trade-off between level-balance and pairwise orthogonality
corresponding to different priors.

\subsection{Supersaturated designs}

A common application of the first-order maximal model is the case of
supersaturated two-level designs where the number of factors is not
less than the number of runs ($m\ge N$) and the first-order maximal
model is not estimable. These designs are popular for screening experiments
with the first-order model -- see \cite{schoen17} for recent developments
of these designs.
%\cite{singh23}

In the context of saturated or supersaturated designs, the most popular
criterion for choosing designs is the
$\text{E}(s^2)$-criterion suggested by \cite{lin93} which is to choose
the design with the smallest $b_2$ among the level-balanced designs
with $b_1=0$.  More recently, \cite{jones14} suggested that there is
no need to impose the restriction of level-balance and introduced
$\text{UE}(s^2)$ supersaturated designs. We note that this criterion
is equivalent to minimising $b_1+b_2$.  The $Q_B$-criterion on the
other hand selects a design depending on $\pi_1$, the prior
probability of the importance of each factor. For $\pi_1 \rightarrow
0$, $Q_B$ reduces to the $\text{E}(s^2)$ criterion, whereas for $\pi_1 =
\frac{1}{2}$, $Q_B$ reduces to $\text{UE}(s^2)$. The study of
$\text{E}(s^2)$ and $\text{UE}(s^2)$-optimal supersaturated designs in
\cite{cheng18} indicated that $\text{E}(s^2)$-optimality is better
when we are interested in models with small number of factors. This
coincides with our results using the $Q_B$-criterion that when $\pi_1$ is
small, designs with more level-balanced factors are recommended.  The
$Q_B$-criterion not only provides a more meaningful way to choose
between the $\text{E}(s^2)$ and $\text{UE}(s^2)$ criteria, it also
provides infinitely many more criteria corresponding to different
values of $\pi_1$.

\begin{example} 

Consider an example of $m=14$ factors and $N=12$ runs.  Table
\ref{table:supersat} gives three supersaturated main effects designs
where $d_1$ is an $\text{E}(s^2)$-optimal design and the other two are
$\text{UE}(s^2)$-optimal designs.  These designs are $d_1, d_2$ and
$d_6$ in Table 1 of \cite{cheng18} but we rearranged the designs to
have the non-level-balanced factors followed by the level-balanced
factors. The values of $(b_1, b_2)$ for these designs are $(0,
\frac{8}{3})$, $(\frac{2}{9}, \frac{19}{9})$, and $(\frac{1}{3}, 2)$,
respectively. We note that $d_1$ is an $\text{E}(s^2)$-optimal design
so all factors are level-balanced and $b_1=0$.  But by fixing the
requirement of level-balance, the aliasing between pairs of factors is
more serious than those of $d_2$ and $d_3$.
 Designs $d_2$ and $d_3$ are $\text{UE}(s^2)$-optimal and both have $b_1+b_2=7/3$.  

%tab1m14.txt 
%QBmain_ex.R
\begin{table}[tb]
\caption{Three supersaturated designs with $m=14$, $N=12$\label{table:supersat}}
{\scriptsize
 \begin{multicols}{3}
 \renewcommand{\arraystretch}{.9}\addtolength{\tabcolsep}{-5pt}
 \begin{tabular}{*{14}{r}}
 \multicolumn{14}{c}{$d_1$}\\\hline
 A & B & C & D & E & F & G & H & I & J & K & L & M & N \\ 
%$x_1$ & $x_2$ & $x_3$ & $x_4$ & $x_5$ & $x_6$ & $x_7$ & $x_8$ & $x_9$ & $x_{10}$ & $x_{11}$ & $x_{12}$ & $x_{13}$ & $x_{14}$ \\ 
 1 & 1 & 1 & 1 & 1 & -1 & 1 & -1 & -1 & 1 & 1 & -1 & 1 & -1 \\ 
 1 & 1 & 1 & -1 & -1 & -1 & 1 & 1 & 1 & -1 & 1 & -1 & -1 & 1 \\
-1 & 1 & -1 & -1 & 1 & 1 & 1 & -1 & 1 & -1 & -1 & -1 & 1 & 1 \\ 
 1 & -1 & 1 & 1 & 1 & -1 & -1 & -1 & -1 & -1 & -1 & 1 & -1 & 1 \\ 
 1 & -1 & -1 & -1 & -1 & -1 & 1 & 1 & -1 & -1 & -1 & 1 & 1 & -1 \\ 
-1 & -1 & 1 & 1 & -1 & 1 & 1 & 1 & -1 & 1 & -1 & -1 & 1 & 1 \\ 
-1 & 1 & -1 & 1 & 1 & 1 & -1 & 1 & -1 & -1 & 1 & -1 & -1 & -1 \\ 
 -1 & -1 & -1 & 1 & 1 & -1 & -1 & 1 & 1 & 1 & 1 & 1 & 1 & 1 \\ 
 -1 & 1 & -1 & 1 & -1 & -1 & 1 & -1 & 1 & 1 & -1 & 1 & -1 & -1 \\ 
 -1 & -1 & 1 & -1 & -1 & 1 & -1 & -1 & 1 & -1 & 1 & 1 & 1 & -1 \\ 
 1 & -1 & 1 & -1 & 1 & 1 & -1 & 1 & 1 & 1 & -1 & -1 & -1 & -1 \\ 
 1 & 1 & -1 & -1 & -1 & 1 & -1 & -1 & -1 & 1 & 1 & 1 & -1 & 1 \\ \hline
\end{tabular}
 \begin{tabular}{*{13}{r}r}
 \multicolumn{14}{c}{$d_2$}\\\hline
 A & B & C & D & E & F & G & H & I & J & K & L & M & N \\ 
1 & 1 & 1 & 1 & 1 & 1 & 1 & 1 & 1 & 1 & 1 & 1 & 1 & 1 \\ 
 1 & 1 & 1 & 1 & -1 & -1 & -1 & -1 & 1 & 1 & -1 & -1 & -1 & -1 \\ 
 1 & 1 & -1 & -1 & 1 & 1 & -1 & -1 & -1 & -1 & 1 & 1 & -1 & -1 \\ 
 1 & 1 & -1 & -1 & -1 & -1 & 1 & 1 & -1 & -1 & -1 & -1 & 1 & 1 \\ 
 -1 & -1 & 1 & 1 & 1 & 1 & -1 & -1 & -1 & -1 & -1 & -1 & 1 & 1 \\ 
 -1 & -1 & 1 & 1 & -1 & -1 & 1 & 1 & -1 & -1 & 1 & 1 & -1 & -1 \\ 
 -1 & -1 & -1 & -1 & 1 & 1 & 1 & 1 & 1 & 1 & -1 & -1 & -1 & -1 \\ 
 -1 & -1 & -1 & -1 & -1 & -1 & -1 & -1 & 1 & 1 & 1 & 1 & 1 & 1 \\ 
 1 & -1 & 1 & -1 & 1 & -1 & 1 & -1 & 1 & -1 & 1 & -1 & 1 & -1 \\ 
 1 & -1 & 1 & -1 & -1 & 1 & -1 & 1 & 1 & -1 & -1 & 1 & -1 & 1 \\ 
 1 & -1 & -1 & 1 & 1 & -1 & -1 & 1 & -1 & 1 & 1 & -1 & -1 & 1 \\ 
 -1 & 1 & 1 & -1 & 1 & -1 & -1 & 1 & -1 & 1 & -1 & 1 & 1 & -1 \\ \hline
\end{tabular}
\begin{tabular}{*{14}{r}}
\multicolumn{14}{c}{$d_3$}\\\hline
 A & B & C & D & E & F & G & H & I & J & K & L & M & N \\ 
 1 & 1 & 1 & 1 & 1 & 1 & 1 & 1 & 1 & 1 & 1 & 1 & 1 & 1 \\ 
 1 & 1 & 1 & -1 & -1 & -1 & -1 & 1 & 1 & 1 & 1 & -1 & -1 & -1 \\ 
 1 & 1 & 1 & 1 & 1 & 1 & 1 & -1 & -1 & -1 & -1 & -1 & -1 & -1 \\ 
 1 & 1 & 1 & -1 & -1 & -1 & -1 & -1 & -1 & -1 & -1 & 1 & 1 & 1 \\ 
 -1 & 1 & -1 & 1 & -1 & 1 & -1 & -1 & -1 & 1 & 1 & 1 & -1 & 1 \\ 
 -1 & 1 & -1 & -1 & 1 & -1 & 1 & -1 & -1 & 1 & 1 & -1 & 1 & -1 \\ 
 -1 & 1 & -1 & 1 & -1 & 1 & -1 & 1 & 1 & -1 & -1 & -1 & 1 & -1 \\ 
 -1 & 1 & -1 & -1 & 1 & -1 & 1 & 1 & 1 & -1 & -1 & 1 & -1 & 1 \\ 
 1 & -1 & -1 & -1 & -1 & 1 & 1 & 1 & -1 & 1 & -1 & 1 & 1 & -1 \\ 
 1 & -1 & -1 & 1 & 1 & -1 & -1 & 1 & -1 & 1 & -1 & -1 & -1 & 1 \\ 
 1 & -1 & -1 & -1 & -1 & 1 & 1 & -1 & 1 & -1 & 1 & -1 & -1 & 1 \\ 
 1 & -1 & -1 & 1 & 1 & -1 & -1 & -1 & 1 & -1 & 1 & 1 & 1 & -1 \\ 
 \hline
\end{tabular}
\end{multicols}}
\end{table}

Figure \ref{fig:supersat} shows the $Q_B$ efficiencies for these three
designs for $\pi_1 \in [0.1, 0.8]$. It can be seen that different
designs will be recommended for different priors. The
$\text{E}(s^2)$-optimal design $d_1$ is the best when the expected
number of active factors is less than 2.8 (i.e.\ $\pi_1 \leq 0.2$),
$d_2$ is optimal when the expected number of active factors is between
2.8 and 7 ($0.2 \leq \pi_1 \leq 0.5)$, and $d_3$ is the best when the
expected number of active factors is at least half of the 14 factors
($\pi_1 \geq 0.5$). This illustrates the much richer information to be
gained from studying $Q_B$-optimality over special cases such as
$\text{E}(s^2)$ and $\text{UE}(s^2)$.

\begin{figure}[hbt]
\begin{center}
\includegraphics[width=10cm, height=7.5cm]{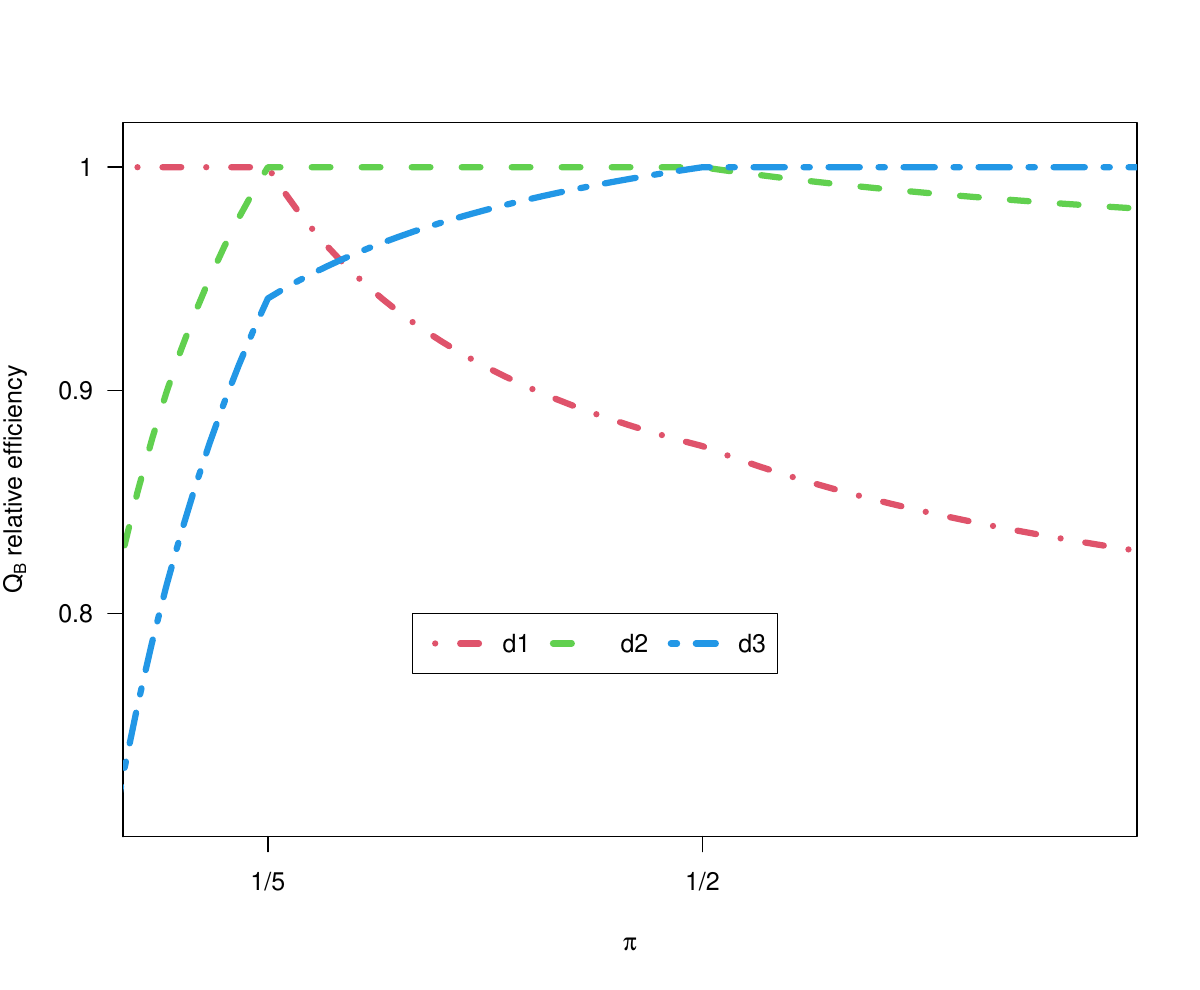}
\caption{Relative $Q_B$-efficiencies for supersaturated designs with
 $m=14$ and $N=12$ for $\pi_1 \in [0.1, 0.8]$\label{fig:supersat}}
\end{center}
\end{figure}
\end{example}

In this paper, we provide a coordinate exchange algorithm to generate
$Q_B$-optimal designs. We are able to generate these designs with the algorithm
and some different $Q_B$-optimal designs can be
found.  For example, for $N=12$, $m=14$ and $\pi_1=0.27$, the algorithm
generate another $Q_B$-optimal design, say $d_4$, which has the same 
values of GWC as those of $d_2$; on the other hand when $\pi_1=0.8$, 
the algorithm generates another $Q_B$-optimal design, say $d_5$, which 
has the same values of GWC as those of $d_3$. Both $d_4$ and $d_5$ are  
$\text{UE}(s^2)$-optimal. The design plans and the $X^\top X$ matrices for 
these designs are in case 1 of the supplementary material. Note that the 
$Q_B$-optimal designs generated by the coordinate exchange algorithm are
either $\text{E}(s^2)$ or $\text{UE}(s^2)$ optimal for the case of $N=12$ 
and $m=14$.

\subsection{Saturated main effect designs}

Saturated designs have run sizes equal to the number of parameters for
the first order maximal model, i.e.\ $m=N-1$. \cite{tsai16} studied a
new class of $Q_B$-optimal saturated main-effect designs without the
requirement of level-balance and provide a novel method of
construction of $Q_B$-optimal saturated main-effect designs by a
modification of conference matrices. The explicit patterns of the
$X^\top X$ matrices for the $Q_B$-optimal designs under different $\pi_1$ are
given.  Often there are several $Q_B$-optimal designs which can be
generated from conference matrices and then the one with the smallest
$A_s$-criterion function value for the full main-effects model is
reported to be the best one.

% There are two types of columns in the saturated main-effect designs: one is
% "level-balanced" where the column has the same number of occurrences
% of 1 and $-1$ and the other is "non-level-balanced" where the number
% occurrences of 1 and $-1$ differ by 2. \cite{tsai16} explicitly show
% the relationship between the number of level-balanced factors (and
% non-level-balanced factors) and the prior probability value $\pi$ for
% $N\equiv 2 (\mod 4)$. \cite{tsai16} show that if $n_1=m$, i.e., all
% factors are level-balanced, then the designs are $Q_B$-optimal for
% $\pi\le 1/(2N-4)$, if $n_1=N/2$, then the designs are $Q_B$-optimal
% for $\pi>1/4$, and if $N/2< n_1<N-1$, then the designs are
% $Q_B$-optimal for $1/(4n_1-2N+4)< \pi\le 1/(4n_1-2N)$.
% In other words, for $N\equiv 2 \mod 4$ and $m=N-1$, we break $\pi$
% into $k=(m+1)/2$ pieces with $0=\alpha_0<\alpha_1<\cdots< \alpha_k=1$
% and $\alpha_i=1/(2m+2-4i)$, for $i=1, \ldots k-1$. Let
% $L_j=(\alpha_{j-1}, \alpha_{j}]$ be the $j$th of the $k$ intervals of
% $\pi$, then designs with $n_1=N-j$ level-balanced factors and $j-1$
% non-level-balanced factors are $Q_B$-optimal for $\pi \in L_j$.

% In \cite{tsai16}, the $Q_B$-optimal saturated main effect designs are
% obtained by the modification of conference matrices by replacing the
% diagonal entries by an appropriate vector of $\pm 1$. 
% However, it is known that the conference matrix does not exist for
% some particular run size, such as N = 22 and sometimes the design
% obtained is not optimal for the main-effect model thus a better
% $Q_B$-optimal design exits. Thus,
Here, we use the coordinate-exchange algorithm discussed in Section 3
to generate $Q_B$-optimal main effects designs and use the
$A_s$-criterion for the main-effect model as the secondary criterion.
Here the $A_s$-efficiency is computed by
$m/(N\mbox{tr}\{(M^{-1})_{22}\})$ where
$(M^{-1})_{22}=(D^{'}Q_0D)^{-1}$, $Q_0={\bf I}-(1/N){\bf 1}{\bf
  1}^{'}$ and $D$ is the model matrix without the column of 1s.

\begin{example}
%{\bf Example 1} \quad
Consider the $Q_B$-optimal saturated main-effects design for the case
with $N=10$ and $m=9$.  \cite{tsai16} show that the $Q_B$-optimal
designs have two types of columns: one has the same
number of $\pm 1$s and the other has the number occurrences of 1 and
$-1$ differ by 2, which is called "non-level-balanced", and the
numbers of level-balanced and non-level-balanced factors depend on the
prior probability of each factor being in the best model, $\pi_1$.
For $N=10$ and $m=9$, the $Q_B$-optimal designs have 9, 8, 7, 6 and 5
level-balanced factors when $\pi_1$ is in each of the intervals
$(0, 1/16]$, $[1/16, 1/12]$, $[1/12, 1/8]$, $[1/8, 1/4]$, $[1/4, 1]$,
respectively.
We can either use the method of the modification of conference
matrices or our exchange algorithm to construct $Q_B$-optimal
saturated main-effects designs for different values of $\pi_1$. Both
methods generate $Q_B$-optimal designs with the correct number of
level-balanced factors. Table \ref{tab:ME1} gives the $Q_B$-optimal
designs generated by these two methods and the corresponding
$A_s$-efficiencies.  Note that to run $m=9$ two-level factors in 10
runs, the theoretical bound with all factors orthogonal to each other
is not achievable, so the low values of the reported efficiencies do
not indicate poor designs. Details of these designs can be found in
case 2 of the supplementary material.
\end{example}

\begin{table}[ht]
 \caption{$A_s$-efficiencies for $Q_B$-optimal saturated main-effect
 designs under different priors for $N=10$}
 \label{tab:ME1}
 \begin{center}
\begin{tabular}{c|*{6}{c}}
 $\pi$& (0, 1/16] & [1/16, 1/12] & [1/12, 1/8] & [1/8, 1/4] & [1/4, 1]
 \\\hline 
$\#$ LB & $9$ & $8$ & $7$ & $6$& $5$ \\\hline

Conference matrix & 0.593 & 0.640 & 0.678 & 0.716& 0.741\\
Algorithm  & 0.659 & 0.685 & 0.689& 0.742 &0.8~~~~\\
\end{tabular}
\end{center}
\end{table}

\subsection{Unsaturated main effects designs}
When $N$ is a power of 2, the
saturated regular fractional factorial designs and any projection to a
subset of columns of these designs are $Q_B$-optimal designs for the
main effects model since in these designs all factors are
level-balanced and all pairs of factors are orthogonal to each
other. When $N$ is a multiple of 4, we look at the Plackett-Burman
designs and their projections. In this section, we extend \cite{tsai16}'s results for saturated main
effects designs to unsaturated main effects designs where $m<N-1$ for
the case when $N \equiv 2 \mod 4$.  

\begin{lemma}
Let $X$ be an $N\times (m+1)$ ($-1, 1$)-matrix where $m\le N-1$, and
$N\equiv 2 \mod 4$. Without loss of generality, suppose that all the
entries in the first column are 1. Consider the class of designs such
that each of the following $m-n_1$ columns has an even number of 1s,
and each of the last $n_1$ columns has an odd number of 1s. When $m$
is odd, $(m+1)/2\le n_1\le m$ and, when $m$ is even, $m/2\le n_1\le
m$. Then if the information matrix of $X$ has the following block
diagonal form
\begin{equation}
 X^tX=\begin{bmatrix}
 (N\pm 2)I_{m+1-n_1}\mp 2J_{m+1-n_1} && 0 \cr
 {0} & 	&	(N\pm 2)I_{n_1}\mp 2J_{n_1}
 \end{bmatrix},
 \label{eq:eq4}
 \end{equation}
$X$ is $Q_B$-optimal for a specific value of $n_1$ within the
class of designs whose entries are all $\pm 1$.
\end{lemma}

\noindent
\begin{proof}
The proof is similar to that of \cite{tsai16}. The $Q_B$-criterion is
the sum of the off-diagonal terms of the information matrix, thus the
off-diagonal blocks of the information matrix should be equal to 0 for
some $n_1$.
\end{proof}

Note that for $N\equiv 2 \mod 4$, the above lemma says that in
$Q_B$-optimal designs, the column having an odd number of 1s is
level-balanced and the column having an even number of 1s has the numbers
of $\pm 1$ differing by 2.
%These two are called level-balanced and non-level-balanced factors.
The value of the $Q_B$-criterion function for a design with the
information matrix for the first-order model with the above block
diagonal form
is 
\begin{equation}
\frac{4\pi_1(m-n_1)+4\pi_1^2[(m-n_1)^2+n_1^2-m]}{N^2}.
\label{eq:qbmain}
\end{equation}
%where $m/2\le n_1\le m$ when $m$ is even and $(m+1)/2\le n_1\le m$
%when $m$ is odd.
The following theorem gives a $Q_B$-optimal design with an appropriate
number of level-balanced factors $(n_1)$ for a given range of $\pi_1$.

\begin{theorem}

For $N\equiv 2 \mod 4$, consider a design with $m-n_1$
non-level-balanced factors and $n_1$ level-balanced factors.
%Let K be a bound on the number of non-level balanced factors
There are $K$ different ranges of $\pi_1$ where the numbers of
level-balanced and non-level-balanced factors in a $Q_B$-optimal
design will change, where $K=(m+1)/2$ when $m$ is odd and $K=m/2+1$
when $m$ is even.  Let $\alpha_0, \cdots, \alpha_K$ be the end points
of each of the $K$ intervals where $\alpha_0=0$,
$\alpha_k=1/(2m+2-4k)$ and $\alpha_K=1$, $k=1, 2, \cdots, K-1$.  Then
for $\pi_1 \in [\alpha_{k-1}, \alpha_{k}]$, $k=1,\cdots, K$, the
  design with $k-1$ non-level-balanced factors and $m-(k-1)$
  level-balanced factors is a $Q_B$-optimal design.
\end{theorem}

\noindent
The proof is simple following equation (\ref{eq:qbmain}).

We see that when $m$ is even, if $\pi_1>1/2$ then designs with the
$m/2$ level-balanced and $m/2$ non-level-balanced factors with the
information matrices of the block-diagonal forms are $Q_B$-optimal;
when $m$ is odd, if $\pi_1>1/4$, designs with the $(m+1)/2$
level-balanced and $(m-1)/2$ non-level-balanced factors with the
information matrices of the block-diagonal forms are $Q_B$-optimal.

\begin{example}
Consider the case of $N=14$ and $m=12$.  According to Theorem 1,
there are $7$ different intervals of $\pi_1$ resulted in 
different $Q_B$-optimal designs, i.e.\ 
$\pi_1\in$ $(0, 1/22]$, $[1/22, 1/18]$, $[1/18,
1/14]$, $[1/14, 1/10]$, $[1/10, 1/6]$, $[1/6, 1/2]$ and $[1/2, 1]$.
For each of the intervals, the $Q_B$-optimal design has 0, 1, 2, 3, 4,
5 and 6 non-level-balanced factors, respectively, with the $X^\top X$ of the block
diagonal forms.

%Similarly, according to Theorem 1 if $N=14$ and $m=11$, the
%$Q_B$-optimal design has 0, 1, 2, 3, 4, 5, and 6 non-level-balanced
%factors when $\pi_1\in $ $(0, 1/20]$, $(1/20, 1/16]$, $(1/16, 1/12]$,
%$(1/12, 1/8]$, $(1/8, 1/4]$, and $(1/4, 1]$.
 \end{example}

Note that the coordinate exchange algorithm sometimes fails to generate
the designs with these specific patterns for large $N$ and $m$.  For example,
%when $N=22$ and $m=14$ the algorithm generates For example, 
when $N=22,
m=15$ and $\pi_1=0.2 \in [\alpha_{6}=1/8, \alpha_{7}=1/4]$, the
algorithm can generate a $Q_B$-optimal design with 6
non-level-balanced factors with the block diagonal pattern, but when
$\pi_1$ is approaching 0, say $0.03 \in (\alpha_{0}=0,
\alpha_{1}=1/28]$, the algorithm generates designs with no
non-level-balanced factor and 15 level-balanced factors but it fails
to have all the off-diagonal elements equal to $\pm 2$.  The details
of these designs for $N=14$ and $N=22$ are in case 3 of the
supplementary material.

\section{Second-order maximal model}
%$E(y)=\beta_0+\beta_1 X_{1}+\beta_2 X_{2}+\cdots +\beta_m
%X_m+\beta_{1,2} X_1X_2 +\beta_{1,3} X_1X_3 +\cdots + \beta_{m-1,
%m}X_{m-1}X_m$ if the second-order main effect model is the maximal
%model. Of course models with other higher-oder effects can be used as
%the the maximal model, but we only discuss these two cases to
%simplyour presentation.
When the second-order model is the maximal model, $\binom{m}{2}$ terms
for the two-factor interaction are added to the model matrix $X$ with
the intercept and $m$ main effects so $v=m+\binom{m}{2}$. Then terms
in the $Q_B$-criterion function in \eqref{eq:QB} can be summarised as
the aliasing between main effects and the intercept, the aliasing
between interactions and the intercept, the aliasing between pairs of
main effects, the aliasing for a main effect and an interaction, and
the aliasing for pairs of interactions, i.e., {\footnotesize
 \begin{equation}
%\sum\limits_{i=1}^{m} \frac{a_{i0}^2 \ p_{i0}}{a^2_{ii}a_{00}}+
 %\sum\limits_{i=1}^{m}\sum\limits_{j=1}^{m}\,
 %\frac{a_{ij}^2\ p_{ij}}{a^2_{ii}a_{jj}} +
 %\sum\limits_{i=m+1}^{v} \frac{a_{i0}^2 \ p_{i0}}{a^2_{ii}a_{00}}+
 %\sum\limits_{i=1}^{m}\sum\limits_{j=m+1}^{v}\,
 %\frac{a_{ij}^2\ p_{ij}}{a^2_{ii}a_{jj}}+
 %\sum\limits_{i=m+1}^{v}\sum\limits_{j=1}^{m}\,
 %\frac{a_{ij}^2\ p_{ij}}{a^2_{ii}a_{jj}}+
 %\sum\limits_{i=m+1}^{v}\sum\limits_{j=m+1}^{v}\,
 %\frac{a_{ij}^2\ p_{ij}}{a^2_{ii}a_{jj}}.
 \sum\limits_{i=1}^{m} \frac{a_{i0}^2 }{N^2} p_{i0}+
 \mathop{\sum\limits_{i=1}^{m}\sum\limits_{j=1}^{m}}_{i\ne j}\,
 \frac{a_{ij}^2}{N^2} p_{ij}+
 \sum\limits_{i=m+1}^{v} \frac{a_{i0}^2 }{N^2}p_{i0}+
 \sum\limits_{i=1}^{m}\sum\limits_{j=m+1}^{v}\,
 \frac{a_{ij}^2}{N^2}\ p_{ij}+
 \sum\limits_{i=m+1}^{v}\sum\limits_{j=1}^{m}\,
 \frac{a_{ij}^2}{N^2}\ p_{ij}+
 \mathop{\sum\limits_{i=m+1}^{v}\sum\limits_{j=m+1}^{v}}_{i\ne j}\,
 \frac{a_{ij}^2}{N^2}\ p_{ij}.
 \label{eq:QB.2nd}
\end{equation}}

We have the following simplifications.
\begin{enumerate}

\item $\sum\limits_{i=1}^{m} {a_{i0}^2}/{N^2}=b_1$.
\item  
 $\underset{i\neq j}{\sum\limits_{i=1}^{m}\sum\limits_{j=1}^m}\,
 {a_{ij}^2}/{N^2}=2b_2$. 
%$\sum\limits_{i=1}^{m}\sum\limits_{\stackrel{j=1}{i \neq j}}^{m}{a_{ij}^2}/{N^2}=2b_2$.
    
\item $\sum\limits_{i=m+1}^{v} {a_{i0}^2}/{N^2}=b_2$ for $i$ referring to an interaction.

\item For the case where $i$ refers to a main effect and $j$ an
  interaction in the fourth term, there are different cases depending on whether or not the main effect and the
  interaction that $i$ and $j$ refer to have a common factor
  or not. For the case with a common factor, say the main effect of
  $X_1$ and the interaction of $X_1$ and $X_2$, we have $X_1\times
  (X_1X_2)=X_1^2X_2=X_2$ and the $a_{ij}$ for this case is the sum of
  element-by-element products of column $X_2$ and a vector of 1s. For the case where $i$ and $j$ refer to a main effect and an
  interaction with no common factor, $a_{ij}$ is the sum of the
  element-by-element products of the corresponding set of three
  columns of $X_d$ which is $R_3(s)$ defined in Section 2. Thus, we
  have $\sum\limits_{i=1}^{m}\sum\limits_{j=m+1}^{v}
  {a_{ij}^2}/{N^2}=(m-1)b_1+3b_3$ since for each of the $m$ factors,
  there are $(m-1)$ interactions involving that factor and there are
  $\binom{m-1}{2}$ interactions with no common factor.
  
\item For the case where $i$ refers to an interaction and $j$ a
  main effect, the details are the same as those for $i$ referring
  to a main effect and $j$ an interaction.

\item For the case where $i$ and $j$ refer to a pair of two
  interactions, again we  discuss two cases depending on whether the pair of
  interactions that $i$ and $j$ refer to have a common factor or
  not. For the case with a common factor, say $X_1X_2$ and $X_1X_3$,
  $a_{ij}$ is the sum of the element-by-element products of columns of
  $X_2$ and $X_3$. For that with no common factor, say
  $X_1X_2$ and $X_3X_4$, $a_{ij}$ is the sum of the
  element-by-element products of the columns corresponding to the four
  factors involved in the pairs of interactions.  Then we have
  $\underset{i\neq j}{\sum\limits_{i=m+1}^{v}\sum\limits_{j=m+1}^v}\,
 {a_{ij}^2}/{N^2}=2(m-2)b_2+6b_4$
%$\sum\limits_{i=m+1}^{v}\sum\limits_{\stackrel{j=m+1}{i\neq j}}^{v}  {a_{ij}^2}/{N^2}=2(m-2)b_2+6b_4$ 
 since for each of the
  $\binom{m}{2}$ interactions there are 
  $2(m-2)$ pairs of interactions with a common factor and there are
  $\binom{m-2}{2}$ interactions with no common
  factor.
 \end{enumerate}
 
For considering sub-models of the second-order maximal model, the
marginality principle of \cite{mccullagh89} is used which means that every term in the model must be accompanied by all terms
marginal to it, whether these are large or small. Thus, if factor
$X_1$ turns out to have a very small main effect, but a large
interaction effect, say $X_1X_2$, then we will still include the main
effect in the model. In screening experiments, it is usually reasonable to assume that factors are
exchangeable, i.e.\ each main effect has the same prior probability $\pi_1$ of being
in the best model and each of the two
interactions has the same the prior probability $\pi_2$ of being in the best
model given that the main effects of both the corresponding factors
are in the model.  Thus, for a model with
main effects of a given set of $a$ factors and $a_2$ two-factor interactions,
the prior probability for this model being the best model is
$\pi_1^{a}(1-\pi_1)^{m-a}\pi_2^{a_2}(1-\pi_2)^{\binom{a}{2}-a_2}$,
where $0\le a_2\le \binom{a}{2}$.  This is used to compute the prior
sum for models being the best where the sum is done over models
containing a given number of main effects (say, $s$) and a given number
of interactions (say $t$). We use $\xi_{st}$ to denote such a prior sum.
Thus
%\cite{tsai10} show that 
for the
second-order maximal model, the $Q_B$-criterion is to select a design that
minimizes
\begin{equation}
Q_B= \xi_{10} b_1 + \xi_{20}( 2b_2) +\xi_{21} b_2+
\xi_{21} \{2(m-1)  b_1\} +\xi_{31}(6 b_3) +
\xi_{32} \{2(m-2) b_2\} +\xi_{42} (6 b_4)\\\nonumber
\label{eq:QBint}
\end{equation}
as given in \cite{tsai10} where the $\xi_{st}$ are computed as follows.

\begin{enumerate}

\item $\xi_{10}$ is the sum of prior probabilities for models
  containing at least a given main effect being the best model, which
  is 
    {\footnotesize
  \[
  \xi_{10} = \pi_1(1-\pi_1)^{m-1}
  +\pi_1\sum\limits_{a=1}^{m-1}\binom{m-1}{a}\pi_1^{a}(
  1-\pi_1)^{m-1-a} \left(\sum\limits_{a_2=0}^{B_1}\binom{B_1}{a_2}
  \pi_2^{a_2}(1-\pi_2)^{B_1-a_2}\right) = \pi_1,
  \]}
%=\pi_1\left[\sum\limits_{a=0}^{m-1}\binom{m-1}{a}\pi_1^a
  % (1-\pi_1)^{m-1-a}\right]=\pi_1$,
 where $B_1=\binom{a+1}{2}$ which is the number of two-factor
 interactions for a set of $a+1$ factors.
 
\item $\xi_{20}$ is the sum of prior probabilities for models
  containing at least a given pair of main effects being the best
  model, which is
    {\footnotesize \[
   \xi_{20} =\pi_1^2 \left[ \sum\limits_{a=0}^{m-2} 
   \binom{m-2}{a} \pi_1^{a} (1-\pi_1)^{m-2-a} \ 
   \left( \sum\limits_{a_2=0}^{B_2} \binom{B_2}{a_2} \pi_2^{a_2}(1-\pi_2)^{B_2-a_2}
   \right) \right] = \pi_1^2,
   \]}
   where
  $B_2=\binom{a+2}{2}$.
    % is number of two-factor interactions for a set of $a+2$ factors.

 \item $\xi_{21}$ is the sum of prior probabilities for models
   containing at least a particular interaction and therefore its
   corresponding main effects, which is 
     {\footnotesize  \begin{align*}
   \xi_{21}  =  & \
   \pi_1^2(1-\pi_1)^{m-2} \pi_2\\
   &+\pi_1^2\pi_2
   \left[ \sum\limits_{a=1}^{m-2} \binom{m-2}{a}
   \pi_1^{a}(1-\pi_1)^{m-2-a}
     \left( \sum\limits_{a_2=0}^{B_2-1} \binom{B_2-1}{a_2}
     \pi_2^{a_2}(1-\pi_2)^{B_2-a_2-1} \right)
     \right]\\
     = &\ \pi_1^2\pi_2.
     \end{align*}}
   % where $B_2=\binom{a+2}{2}$ is the number of two-factor interactions
   % for $a+2$ factors. 
   This corresponds to the aliasing for the case when $i$ refers
   to an interaction and $j$ is the intercept as well as to the case
   when $i$ and $j$ refer to a main effect and an interaction
   with a common factor.

\item $\xi_{31}$ is the sum of prior probabilities for models
  containing at least 3 main effects and an interaction of these
  factors, which is 
    {\footnotesize
  \[
  \xi_{31}=
  \pi_1^3 \pi_2 \left[ 
  \sum\limits_{a=0}^{m-3} 
  \binom{m-3}{a} \pi_1^{a} (1-\pi_1)^{m-3-a}
  \left( 
  \sum\limits_{a_2=0}^{B_3-1} \binom{B_3-1}{a_2} \pi_2^{a_2} (1-\pi_2)^{B_3-a_2-1}
  \right)
    \right]=\pi_1^3\pi_2,
    \]}
  where $B_3 =\binom{a+3}{2}$. % is number of two-factor interactions
                               % for a set of $a+3$ factors.
This corresponds to the aliasing for the case when $i$ and $j$ refer to a main effect and an interaction with no common factor.

\item $\xi_{32}$ is the sum of prior probabilities for models
  containing at least main effects of a given 3 factors and two
  interactions of these factors, which is
   {\footnotesize \[
\xi_{32}=\pi_1^3\pi_2^2\left[\sum\limits_{a=0}^{m-3} \binom{m-3}{a}\pi_1^{a}(
    1-\pi_1)^{m-3-a} \left(\sum\limits_{a_2=0}^{B_3-2} \binom{B_3-2}{a_2}
    \pi_2^{a_2}(1-\pi_2)^{B_3-a_2-2} \right) \right]=\pi_1^3\pi_2^2.
    \]}

    \item $\xi_{42}$ is the sum of prior probabilities for models
      containing at least main effects of a given 4 factors and an
      interaction of these factors, which is  
        {\footnotesize \[
      \xi_{42}= \pi_1^4 \pi_2^2
      \left[ \sum\limits_{a=0}^{m-4} \binom{m-4}{a}\pi_1^{a} (
        1-\pi_1)^{m-4-a}
        \left( \sum\limits_{a_2=0}^{B_4-2} \binom{B_4-2}{a_2}
        \pi_2^{a_2}(1-\pi_2)^{B_4-a_2-2} \right)
        \right]=\pi_1^4\pi_2^2,
        \]}
        where $B_4=\binom{a+4}{2}$.

  %is number of two-factor interactions for a set of $a+4$ factors.
 \end{enumerate}

It follows that the $Q_B$-criterion for the second-order maximal model is 
 \begin{equation}
Q_B=\left\{\pi_1+2(m-1)\pi_1^2\pi_2\right\} b_1+
\left\{2\pi_1^2+\pi_1^2\pi_2+2(m-2)\pi_1^3\pi_2^2\right \} b_2+
6\pi_1^3\pi_2 b_3+6\pi_1^4\pi_2^2 b_4.
\label{eq:QB2nd2}
\end{equation}
This is a linear function of the generalized word counts $b_1$, $b_2$,
$b_3$ and $b_4$ with weights depending on the prior knowledge specified by
$\pi_1$ and $\pi_2$. 
Note that here we use marginality to defined the class of possible models
and $\pi_1$ is the prior probability that a main effect is in the best model.
\cite{mee17} modified the $Q_B$-criterion to the case where effect heredity 
is used, but marginality is not.

In the following section, we will demonstrate the use of this criterion as the primary objective to generate a wide range
of second-order $Q_B$-optimal designs without the requirements of level-balance and pairwise orthogonality.

\begin{example}
%{\bf Example 4}\quad 

For the second-order $Q_B$-optimal designs, we consider the case of
$N=12$ and $m=4$. We discuss two designs: one is a  submatrix of
 the Hadamard matrix and the other
 is  generated by our algorithm with $\pi_1=0.8$
and $\pi_2=0.8$; both designs are given in Table \ref{tab:N12m4}. The first design is a level-balanced design with
$(b_1, b_2, b_3, b_4)= (0, 0, 4/9, 1/9)$ and the second one is a
non-level-balanced design with $(b_1, b_2, b_3, b_4)=(1/9, 0, 1/9,
1/9)$ where all the main effects are partially aliased with the
intercept. The aliasing patterns between main effects and interactions
in the second design are less serious than those in the first
design. In terms of the usual minimum aberration criterion, the first
one is a better design, but in terms of $Q_B$ as in equation
\eqref{eq:QB2nd2}, the second design will be recommended if models
with more parameters are of interest. For example, if we set
$\pi_1=0.8$, the second design has lower $Q_B$-value when $\pi_2>0.1$.
%Details of these designs are in case 5 of the supplementary material. 

\begin{table}[thbp]
\centering
\caption{12-run designs with four two-level factors}
\label{tab:N12m4}
\begin{footnotesize}

\begin{multicols}{2}
\renewcommand{\arraystretch}{.6}\addtolength{\tabcolsep}{0pt}
\begin{tabular}{rrrr}
x1 & x2 & x3 & x4 \\ \hline
1 & 1 & 1 & 1 \\ 
 -1 & -1 & 1 & -1 \\ 
 -1 & -1 & -1 & 1 \\ 
 1 & -1 & -1 & -1 \\ 
 1 & 1 & -1 & -1 \\ 
 1 & 1 & 1 & -1 \\ 
 -1 & 1 & 1 & 1 \\ 
 1 & -1 & 1 & 1 \\ 
 -1 & 1 & -1 & 1 \\ 
 -1 & -1 & 1 & -1 \\ 
 1 & -1 & -1 & 1 \\ 
 -1 & 1 & -1 & -1 \\ \hline
\end{tabular}

\begin{tabular}{rrrr}
x1 & x2 & x3 & x4 \\ \hline
1 & 1 & -1 & -1 \\ 
 -1 & -1 & 1 & 1 \\ 
 -1 & 1 & 1 & 1 \\ 
 -1 & -1 & -1 & -1 \\ 
 1 & 1 & -1 & 1 \\ 
 1 & -1 & 1 & -1 \\ 
 -1 & 1 & -1 & 1 \\ 
 -1 & -1 & -1 & 1 \\ 
 -1 & 1 & 1 & -1 \\ 
 -1 & 1 & -1 & -1 \\ 
 1 & 1 & 1 & 1 \\ 
 1 & -1 & -1 & 1 \\ \hline
\end{tabular}
\end{multicols}
\begin{multicols}{2}
\begin{verbatim}
       A  B  C  D AB AC AD BC BD CD
   12  0  0  0  0  0  0  0  0  0  0
A   0 12  0  0  0  0  0  0  4 -4  4
B   0  0 12  0  0  0  4 -4  0  0  4
C   0  0  0 12  0  4  0  4  0  4  0
D   0  0  0  0 12 -4  4  0  4  0  0
AB  0  0  0  4 -4 12  0  0  0  0 -4
AC  0  0  4  0  4  0 12  0  0 -4  0
AD  0  0 -4  4  0  0  0 12 -4  0  0
BC  0  4  0  0  4  0  0 -4 12  0  0
BD  0 -4  0  4  0  0 -4  0  0 12  0
CD  0  4  4  0  0 -4  0  0  0  0 12

     A  B  C  D AB AC AD BC BD CD
   12 -2  2 -2  2  0  0  0  0  0  0
A  -2 12  0  0  0  2 -2  2 -2  2 -2
B   2  0 12  0  0 -2 -2  2 -2  2  2
C  -2  0  0 12  0 -2 -2 -2  2  2  2
D   2  0  0  0 12  2 -2 -2  2  2 -2
AB  0  2 -2 -2  2 12  0  0  0  0  4
AC  0 -2 -2 -2 -2  0 12  0  0  4  0
AD  0  2  2 -2 -2  0  0 12  4  0  0
BC  0 -2 -2  2  2  0  0  4 12  0  0
BD  0  2  2  2  2  0  4  0  0 12  0
CD  0 -2  2  2 -2  4  0  0  0  0 12
\end{verbatim}
\end{multicols}
\end{footnotesize}
\end{table}

\end{example}

Most work in the design literature focuses on orthogonal main effects
designs with $b_1=b_2=0$. In this case, the $Q_B$-criterion can be
used to select the best second-order design among the class of
orthogonal main-effects designs when the estimation of two-factor
interactions is of interest.  The $Q_B$-criterion looks at the
weighted average of $b_3$ and $b_4$ with weights depending on the
prior probabilities $\pi_1$ and $\pi_2$.
%In this paper we will study a broader class of designs using
 %$Q_B$-criterion as the primary criterion to construct projective
 %two-level designs without the restriction of level balance and
 %orthogonality.  Additionally, we will show that we can use this as a
 %refinement of the $E(s^2)$-criterion among those designs whose
 %$E(s^2)$ are the same in the context of saturated or supersaturated
 %designs in the next section.

\begin{example}

We consider the case with six two-level factors in $N=16$ runs. We
first look at designs obtained from sub-columns of the 16-run Hadamard
matrix given in the supplementary material, from which we have removed the first
column of 1s. There are five classes of orthogonal main-effect designs
for $m=6$ obtained from projections of the Hadamard matrix. The values
of the generalized word counts $b_3$ and $b_4$ for these five
orthogonal main-effects designs are (0,3), (1/2, 2), (1,1), (5/4, 3/4)
and (2,1).
%, where the first, third and the fifth designs are regular designs
% and the other two are irregular designs.
In terms of the $Q_B$-criterion, minimizing $b_3+\pi_1\pi_2 b_4$, the
fourth and fifth designs are not better than the third design over all
possible $\pi_1$ and $\pi_2$. Thus they are not admissible. Also the
second design is optimal only for $q = \pi_1\pi_2 = 1/2$, since when
$0\le q<1/2$, the first design is better than it and when $1/2<q\le 1$
the third design is better than it. So, only the first and third
designs, columns (1 2 4 8 11 13) and (1 2 3 4 8 13) 
%of Table \ref{tab:had16}, 
respectively, are worth studying further. % They are both regular designs. The former has $b_3=0$ and $b_4=3$, and the latter has $b_3=b_4=1$. 

Additionally, the coordinate exchange algorithm is used to generate $Q_B$-optimal designs for the second-order maximal model. $d_1$ and $d_3$ are found using the algorithm when $\pi_1=0.7$ and $\pi_2=0.5$ and when $\pi_1=0.9$ and $\pi_2=0.8$, respectively. For  $\pi_1=0.9$ and $\pi_2=0.8$, the algorithm sometimes generates 
an alternative $Q_B$-optimal design, denoted as $d_6$, which is an irregular design but has the same GWC as that of $d_3$. 
Details of these designs are given in case 4 of the supplementary material.  
We note that in terms of the generalized minimum aberration criterion which minimizes $b_3$ and $b_4$ sequentially, $d_1$ is always the best design. In terms of $Q_B$-values, we see that when models with fewer parameters are of interest, say $\pi_1=0.7$ and $\pi_2=0.5$ and $q<1/2$, $d_1$ is indeed a better design; when models with more parameters are of interest, say $\pi_1=0.9$ and $\pi_2=0.8$ and $q>1/2$, $d_3$ and $d_6$ have lower values of $Q_B$ than $d_1$.

%$q<1/2$, i.e.\ when models with fewer parameters are of interest, say $\pi_1=\pi_2=0.4$, the first design is indeed a better design. However, when models with more parameters are of interest, say $\pi_1=0.9$ and $\pi_2=0.8$, the third design is a better design. 

To have a better understanding of the
properties of these three designs, we report the overall
$A_s(f)$-efficiencies for models with main effects of $f$ factors and
various numbers of interactions and the number of non-estimable models
(NoEst) for all possible $f$-factor projections of these designs, for
$f=3,4,5,6$ in Table \ref{tab:As}.
This table also gives the average of the $A_s$-efficiency for each of
the projections.
It shows that for $f=3$, $d_1$ is better than $d_3$ and $d_6$ since
all its three-factor projections are a replicated $2^3$ full
factorial, while $d_3$ and $d_6$ have $b_3=1$. This coincides with the
conclusions of $Q_B$ that when models with fewer parameters are of
interest, $d_1$ is a better design. When we project onto 4 factors,
$d_1$ is still better than $d_3$ in terms of $A_s$ and NoEst, but it
is worse than $d_6$ for models with all the main effects and more than
4 interactions. When we project onto five factors, if the number of
interactions is 8 or higher, all the possible models from $d_1$ are
not estimable, but there are some models which are estimable in $d_3$
and $d_6$. Also $d_6$ is better than $d_1$ in terms of NoEst. Similar
patterns are observed for models with 6 main effects and some
interactions. 
% We note that there are partial aliasing among main effects and
% interaction, so the number of non-estimable models is less than
% those of the other two designs.

\begin{table}[htbp]
\centering
{\footnotesize 
\caption{Projection properties of three designs with 6 two-level
 factors in 16 runs}
\label{tab:As}
\renewcommand{\arraystretch}{.8}\addtolength{\tabcolsep}{-2pt}
\begin{tabular}{c|c|cr|cr|cr}
&& \multicolumn{2}{c|}{$d_1$}
 & \multicolumn{2}{c|}{$d_3$}
& \multicolumn{2}{c}{$d_6$}\\ \hline

&\# interactions & $A_s$ & NoEst & $A_s$ & NoEst & $A_s$ & NoEst \\ 
 \hline
 $f=3$ 
& 1 & 1.000 & 0 & 0.950 & 3 & 0.971 & 0 \\ 
& 2 & 1.000 & 0 & 0.950 & 3 & 0.958 & 0 \\ 
& 3 & 1.000 & 0 & 0.950 & 1 & 0.950 & 0 \\ 
 \hline
 $f=4$
&1 & 1.000 & 0 & 0.900 & 9 & 0.951 & 0 \\ 
& 2 & 0.960 & 9 & 0.827 & 39 & 0.911 & 0 \\ 
& 3 & 0.880 & 36 & 0.770 & 69 & 0.877 & 0 \\ 
& 4 & 0.800 & 45 & 0.733 & 60 & 0.847 & 0 \\ 
& 5 & 0.800 & 18 & 0.733 & 24 & 0.821 & 0 \\ 
& 6 & 0.800 & 3 & 0.733 & 4 & 0.800 & 0 \\ \hline
$f=5$ 
& 1 & 1.000 & 0 & 0.850 & 9 & 0.933 & 0 \\ 
& 2 & 0.933 & 18 & 0.711 & 78 & 0.872 & 0 \\ 
& 3 & 0.800 & 144 & 0.579 & 303 & 0.811 & 4 \\ 
& 4 & 0.614 & 486 & 0.455 & 687 & 0.750 & 29 \\ 
& 5 & 0.405 & 900 & 0.343 & 993 & 0.684 & 90 \\ 
& 6 & 0.210 & 996 & 0.253 & 941 & 0.611 & 155 \\ 
& 7 & 0.067 & 672 & 0.193 & 581 & 0.527 & 160 \\ 
& 8 & 0.000 & 270 & 0.167 & 225 & 0.428 & 99 \\ 
& 9 & 0.000 & 60 & 0.167 & 50 & 0.309 & 34 \\ 
& 10 & 0.000 & 6 & 0.167 & 5 & 0.167 & 5 \\ \hline
$f=6$
& 1 & 1.000 & 0 & 0.800 & 3 & 0.918 & 0 \\ 
& 2 & 0.914 & 9 & 0.600 & 42 & 0.835 & 0 \\ 
& 3 & 0.747 & 115 & 0.418 & 265 & 0.745 & 10 \\ 
& 4 & 0.527 & 645 & 0.266 & 1002 & 0.641 & 115 \\ 
& 5 & 0.304 & 2091 & 0.152 & 2547 & 0.517 & 603 \\ 
& 6 & 0.128 & 4365 & 0.075 & 4628 & 0.378 & 1873 \\ 
& 7 & 0.030 & 6243 & 0.031 & 6237 & 0.235 & 3775 \\ 
& 8 & 0.000 & 6435 & 0.009 & 6375 & 0.111 & 5115 \\ 
& 9 & 0.000 & 5005 & 0.002 & 4997 & 0.030 & 4717 \\
\end{tabular}}
\end{table}

\end{example}

We also discuss two designs generate by the coordinate exchange algorithm for $N=24$ and $m=7$ in the supplementary material. 
These two designs are both orthogonal main 
effects plan with $b_1=b_2=0$. The first design has $b_3=0$ and $b_4=35/9$ and the second design has $b_3=2/3$ and $b_4=5/3$. Again, in terms of aberration, the first design is always the better design but if
models with more parameters are of interest, the second design would be recommended. 
%Note that for different values of $\pi_1$ and $\pi_2$, designs were generated which are not orthogonal main effects plans.

\section{Discussion}
In this paper, the applications of using the $Q_B$-criterion as the
primary objective for two level designs are given. We demonstrate
that by relaxing the requirements of level-balance and pairwise
orthogonality, a wider range of designs can be recommended. If experimenters are interested in models with
more parameters, then it would be better to go beyond the traditional
$E(s^2)$-designs or the aberration-type criteria. The flexibility of $Q_B$ makes it an appropriate criterion for two-level screening designs in almost all situations. 

\section*{Supplementary Materials}

The supplementary materials contain details of the $Q_B$ designs constructed by the exchange algorithm for different cases discussed in the paper.
\begin{enumerate}[{Case} 1:]\itemsep=0pt
\item Supersaturated designs with $m=14$ and $N=12$.
\item  Saturated main-effects designs with $m=9$ and $N=10$.
\item Unsaturated main-effects designs  with  $m = 12$ and $N = 14$.
\item Three second-order $Q_B$-optimal designs with $m=6$ and $N=16$.
\item Two second-order $Q_B$-optimal designs  designs with $m=7$ and $N=24$.
\end{enumerate}

\par
%%%%%%%%%%%%%%%%%%%%%%%%%%%%%%%%%%%%%%%%%%%%%%%%%%%%%%%%%%%%%%%%%%%%%%%%%%%%%%%%%%%%%%%%%%%%%%%%%%%%%%%%%%%%%%%%%%%%%%%%%%%%

%\section*{Acknowledgements}

%Write the acknowledgements here.
\par

%%%%%%%%%%%%%%%%%%%%%%%%%%%%%%%%%%%%%%%%%%%%%%%%%%%%%%%%%%%%%%%%%%%%%%%%%%%%%%%%%%%%%%%%%%

\bibhang=1.7pc
\bibsep=2pt
\fontsize{9}{14pt plus.8pt minus .6pt}\selectfont
\renewcommand\bibname{\large \bf References}
%\begin{thebibliography}{11}
\expandafter\ifx\csname
natexlab\endcsname\relax\def\natexlab#1{#1}\fi
\expandafter\ifx\csname url\endcsname\relax
  \def\url#1{\texttt{#1}}\fi
\expandafter\ifx\csname urlprefix\endcsname\relax\def\urlprefix{URL}\fi


\begin{thebibliography}{}

\bibitem[\protect\citeauthoryear{Cheng et~al.}{2018}]{cheng18}

  Cheng, C.-S., Das, A., Singh, R. and Tsai, P.-W.  (2018), E($s^2$)-
  and UE($s^2$)-optimal supersaturated designs, {\em Journal of
    Statistical Planning and Inference} {\bf 196},~105--114.
  %\url{https://www.sciencedirect.com/science/article/pii/S0378375817301969}

\bibitem[\protect\citeauthoryear{Fries and Hunter}{1980}]{fries80}
  Fries, A. and Hunter, W.~G. (1980), Minimum aberration $2^{k-p}$
  designs, {\em Technometrics} {\bf 22}(4),~601--608.
%\url{http://www.jstor.org/stable/1268198}

  
\bibitem[\protect\citeauthoryear{Jones and Majumdar}{2014}]{jones14}
  Jones, B. and Majumdar, D.  (2014), Optimal supersaturated
  designs, {\em Journal of the American Statistical Association} {\bf
    109}(508),~1592--1600.
%\url{https://doi.org/10.1080/01621459.2014.938810}

\bibitem[\protect\citeauthoryear{Lin}{1993}]{lin93}
  Lin, D. K.~J.  (1993), A new class of supersaturated designs, {\em
    Technometrics} {\bf 35}(1),~28--31.
%\url{http://www.jstor.org/stable/1269286}

\bibitem[\protect\citeauthoryear{McCullagh and Nelder}{1989}]{mccullagh89}
  McCullagh, P. and Nelder, J.~A.  (1989), {\em Generalized Linear
    Models}, Chapman \& Hall / CRC, London.

\bibitem[\protect\citeauthoryear{Mee et~al.}{2017}]{mee17}
Mee, R.~W., Schoen, E.~D., and Edwards, D.~J. (2017), Selecting an orthogonal or 
nonorthogonal two-level design for screening. {\em Technometrics}, 
{\bf 59}(3),~305--318.
%https://doi.org/10.1080/00401706.2016.1186562



\bibitem[\protect\citeauthoryear{Meyer and Nachtsheim}{1995}]{meyer95}
  Meyer, R.~K. and Nachtsheim, C.~J.  (1995), The coordinate-exchange
  algorithm for constructing exact optimal experimental designs, {\em
    Technometrics} {\bf 37}(1),~60--69.
%\newline\harvardurl{http://www.jstor.org/stable/1269153}

\bibitem[\protect\citeauthoryear{Schoen et~al.}{2017}]{schoen17}
  Schoen, E.~D., Vo-Thanh, N. and Goos, P.  (2017), Two-level
  orthogonal screening designs with 24, 28, 32, and 36 runs, {\em
    Journal of the American Statistical Association} {\bf
    112}(519),~1354--1369.
%\url{http://www.jstor.org/stable/45027982}

%\bibitem[\protect\citeauthoryear{Singh and Stufken}{2023}]{singh23}
%  Singh, R. and Stufken, J.  (2023), Selection of two-level
%  supersaturated designs for main effects models, {\em Technometrics}
%  {\bf 65}(1),~96--104.
%\url{https://doi.org/10.1080/00401706.2022.2102080}

\bibitem[\protect\citeauthoryear{Tang and Deng}{1999}]{tang99}
  Tang, B. and Deng, L.~Y.  (1999), Minimum $G_{2}$-aberration for
  non-regular fractional factorial designs, {\em Annals of
    Statistics} {\bf 27}(4),~1914--1926.
    
\bibitem[\protect\citeauthoryear{Tang}{2001}]{tang01}
 Tang, B.  (2001), Theory of $J$-characteristics for fractional
  factorial designs and projection justification of minimum
  $g_2$‐aberration, {\em Biometrika} {\bf 88}(2),~401--407.
%\url{https://doi.org/10.1093/biomet/88.2.401}


\bibitem[\protect\citeauthoryear{Tsai and Gilmour}{2010}]{tsai10}
  Tsai, P.-W. and Gilmour, S.~G.  (2010), A general criterion for
  factorial designs under model uncertainty, {\em Technometrics} {\bf
    52}(2),~231--242.
%\newline\harvardurl{https://doi.org/10.1198/TECH.2010.08093}

\bibitem[\protect\citeauthoryear{Tsai and Gilmour}{2016}]{tsai16}
  Tsai, P.-W. and Gilmour, S.~G.  (2016), New families of
  $Q_B$-optimal saturated two-level main effects screening designs,
  {\em Statistica Sinica} {\bf 26}(2),~605--617.
%\url{https://www.jstor.org/stable/24721290}

\bibitem[\protect\citeauthoryear{Tsai et~al.}{2007}]{tsai07}
  Tsai, P.-W., Gilmour, S.~G. and Mead, R.  (2007), Three-level
  main-effects designs exploiting prior information about model
  uncertainty, {\em Journal of Statistical Planning and Inference}
  {\bf 137}(2),~619--627.
%\url{https://www.sciencedirect.com/science/article/pii/S0378375806000309}


\bibitem[\protect\citeauthoryear{Vazquez et~al.}{2023}]{vazquez23}
Vazquez, A.~R., Wong, W.~K. and Goos, P. (2023), Constructing two-level $Q_B$-optimal screening designs using mixed-integer programming and heuristic algorithms. {\em Statistics and Computing} {\bf 33}(7)~\texttt{https://doi.org/10.1007/s11222-022-10168-1}. 


\end{thebibliography}
\end{document}